\documentclass{Interspeech2024}
\usepackage{multirow}
\usepackage{adjustbox}
\usepackage{hyperref}

\interspeechcameraready

\graphicspath{{./fig/}}
\title{On Improving Error Resilience of Neural End-to-End Speech Coders}
\name[affiliation={}]{Kishan}{Gupta}
\name[affiliation={}]{Nicola}{Pia}
\name[affiliation={}]{Srikanth}{Korse}
\name[affiliation={}]{Andreas}{Brendel}
\name[affiliation={}]{Guillaume}{Fuchs}
\name[affiliation={}]{Markus}{Multrus}

\address{
  Fraunhofer IIS, Erlangen, Germany.}
\email{kishan.gupta@iis.fraunhofer.de}


\begin{document}

\maketitle

\begin{abstract}
    
    Error resilient tools like Packet Loss Concealment (PLC) and Forward Error Correction (FEC) are essential to maintain a reliable speech communication for applications like Voice over Internet Protocol (VoIP), where packets are frequently delayed and lost. In recent times, end-to-end neural speech codecs have seen a significant rise, due to their ability to transmit speech signal at low bitrates but few considerations were made about their error resilience in a real system. Recently introduced Neural End-to-End Speech Codec (NESC) can reproduce high quality natural speech at low bitrates. We extend its robustness to packet losses by adding a low complexity network to predict the codebook indices in latent space. Furthermore, we propose a method to add an in-band FEC at an additional bitrate of 0.8 kbps. Both subjective and objective assessment indicate the effectiveness of proposed methods, and demonstrate that coupling PLC and FEC provide significant robustness against packet losses.

\end{abstract}

\section{Introduction}

Voice over Internet Protocol~(VoIP) is the most widely used application in modern digital communication systems. In order to ensure real-time communication, VoIP uses the User Datagram Protocol (UDP) in conjunction with the Real-Time Transport Protocol (RTP) to send encoded audio packets over the network~\cite{VOIP}. Since UDP is an unguaranteed connectionless protocol, the transmission is prone to delay and jitter (delay variation) in packet arrival, and even to packet losses. Modern communication codecs must be capable of handling such packet delays and losses in order to maintain good quality of service. Basic PLC~\cite{BasicPLC} techniques includes methods like silencing the lost frame, repeating the past received frame or some form of time-scaling. Such methods are not very effective and produces audible artefacts. The transmission jitter is generally compensated by a Jitter Buffer Management (JBM)~\cite{EVS_JBM} at the receiver side that can handle out-of-order packets and maintain a steady rate of playback.

More advanced state-of-the-art communication codecs like 3GPP Enhanced Voice Service (EVS)~\cite{EVS_2014alt} support two types of error resilient tools. The first type of tools is the PLC~\cite{EVS_PLC}, which extrapolates coded parameters such as line spectral frequencies (LSF) from previous frames and can also be guided by additionally transmitted parameters.. The other type is an in-band FEC~\cite{EVS_CA} where information of distant past frames are summarily coded and the generated additional information is piggy-backed on the primary payload of future frames. When the current frame is declared as lost during the decoding process, a JBM can exploit the in-band FEC, where a future frame containing redundant information of the current frame might be available in the buffer. Transmitting redundant information in anticipation of a loss has to be done with care usually in a channel-aware mode since it puts an additional strain on a network connection and could engender additional latency. 

In recent times, Deep Neural Network (DNN)-based solutions have been shown to outperform conventional PLC methods for large bursts and high error rates. In the earliest DNN-based solution for concealment~\cite{stft_plc}, a small network of fully connected layers estimates the log power spectrum and the phase of the lost frame. The DNN-based PLC solutions are mostly predictive in nature as they aim to estimate or generate the lost frames based on available past frames. Thus, autoregressive networks are widely used for PLC. In ~\cite{arplc_simple}, a Recurrent Neural Network (RNN)-based network is trained to predict the samples of the next frame given past samples as input. The network has limited concealment capability as prediction error of samples may accumulate quickly over multiple frames and will not be effective for burst losses. Another approach which is based on WaveRNN~\cite{NetEQ}, uses a conditioning network that takes a mel-spectrogram as an input and conditions an autoregressive network to generates the samples. A more powerful method in~\cite{lpcnet_plc} uses a predictive network along with an autoregressive LPCNet vocoder. The predictive network estimates the features of lost frames which are then used to condition LPCNet to generate the missing samples. 

The powerful generation capability of Generative Adversarial Networks (GANs) has also been explored for PLC and in most cases outperforms autoregressive methods. GANs generally employ a generative network that can produce an entire lost frame in one forward pass and are trained adversarially with multiple discriminators representing a trainable loss function. These networks can generate the lost frame either in time domain~\cite{PLAE, TMGAN, liu22s_interspeech, li22ea_interspeech} or in time-frequency domain~\cite{stft_gan}.

All the aforementioned networks generally work as post-processor in combination with conventional or neural speech codecs. Such systems require further processing like cross-fading, overlap-add etc. to ensure seamless transition from decoded to concealed frame and vice-versa. With the advent of end-to-end self-supervised neural speech codecs like~\cite{Soundstream, NESC, Encodec, DAC}, there is a need for more integrated error resilient tools for concealment. The common architecture of end-to-end codecs includes an encoder, a decoder and a Vector Quantizer (VQ) consisting of multiple residual stages to calculate a quantized representation of the encoder output, i.e., the latent. Some approaches have been developed to perform PLC using lost latent representations~\cite{TFNet, FD_plc}. In TFnet codec~\cite{TFNet}, the latent is masked to indicate lost frame and is recovered by either using an additional module after the decoder or with an optimized decoder capable of handling such frame  losses. In~\cite{FD_plc}, an additional block called FD-PLC is inserted between encoder and decoder, and is trained end-to-end to recover the lost quantized features. For in-band FEC, a neural network based solution has been developed for conventional codec~\cite{Valin2023} but to the best of our knowledge no work have been done so far for neural codecs.  

In this paper, we propose a method to perform PLC and in-band FEC in the latent domain for neural end-to-end speech coders like NESC. Our low complexity PLC model uses the quantized latent of the past frame and predicts the likelihood of the next codebook indices. Codebook indices prediction has been proposed previously for entropy coding~\cite{Encodec} or to predict fine residual codebooks~\cite{LMCodec} but not for PLC. Moreover, networks used for these predictions are highly complex language models. Our contribution in this paper can be summarized as below:

\begin{itemize}
	\item {} We propose a causal, convolutional, lightweight model trained to predict future codebook indices. During inference, the model can run auto-regressively to conceal burst losses.
	\item {} We propose to distill sum of multiple code-vectors of residual VQ onto a single low bitrate codebook and use it for concealment.
	\item {} We propose a in-band FEC method for NESC that piggy-back the low bitrate codebook with a future frame. Our solution only adds 0.8 kbps of additional bitrate and do not use neural network, thus it does not introduce any complexity overhead. 
	\item {} Our proposed method is trained independently of the codec and does not require fine-tuning or re-training of the codec. Also, it does not require any extra information regarding the occurrence of packet losses during the decoding process. The predicted and the distilled code-vectors are directly used as input to the neural decoder.
\end{itemize}

\section{Proposed Methods}

\subsection{NESC} 
NESC~\cite{NESC} is an adversarially trained end-to-end neural codec designed to generate good quality speech signals at low bitrates. It consists of a neural encoder, a neural decoder and a learned quantization layer. The encoder operates on a frame size of 10ms with an additional 5 ms of lookahead and 5 ms of past samples. It produces a learned latent representation, for each frame. The latent vector is then quantized with a residual VQ that learns multiple codebooks where the code-vectors in each subsequent codebook quantizes the residual from the previous ones. The output of the quantizer is a sum of one or more code-vectors. The paper~\cite{NESC} proposes to use 3 codebooks, each with 1024 code-vectors, thus, the codec operates at bitrates ranging from 1 to 3 kbps. For our implementation, we train a new NESC model that quantizes latent vector with 4 codebooks, each with 256 code-vectors. We found that this setup increases the quality of the codec and provides more scalability as the operating bitrate now ranges from 0.8 to 3.2 kbps.

\subsection{Codebook Distillation}
In our proposed model for PLC, we predict the codebook indices of a lost frame using past latent code-vectors. Because of multiple codebooks used during quantization, the prediction of all the indices requires multiple models that in-turn increases the complexity overhead. An obvious choice would be to use only the first codebook of NESC but this only provides a low quality concealment and does not model all variations in the speech signals. Hence, we propose a distillation method where a new single codebook is learned using the sum of code-vectors from multiple codebooks. We choose to distill the information from the first two codebooks (1.6 kbps) of NESC onto another "distilled codebook" with 256 code-vectors (0.8kbps). The choice of using only first two codebooks is motivated by the trade-off between achievable quality and effective distillation given a target codebook size. The distilled codebook is used for FEC as low bitrate redundant information as well as for PLC.

\subsection{PLC}
For concealment, a causal convolutional model predicts the newly trained distilled codebook index from the past code-vectors. The PLC model takes the code-vectors of last seven frames \{$C'(n-1), C'(n-2),..,C'(n-7)$\} from the distilled codebook as an input and outputs the conditional distribution of the codebook index $c'(n)$ of the current frame.

\begin{equation}\label{probability}
	\textit{P}_{c'}(c'(n) | C'(n-1), C'(n-2),....,C'(n-7)).
\end{equation}
The architecture of the proposed model contains a 1-D convolutional layer with kernel size = 7, followed by two 1-by-1 convolutional layer with kernel size = 1. Finally, the output is passed through a fully connected layer with 256 hidden units and a softmax layer that outputs the probability distribution over the possible 256 codebook indices. We use LeakyRelu activation after each convolutional layer.

\subsection{FEC}
\begin{figure}[t]
	\centering
	\includegraphics[width=0.8\linewidth]{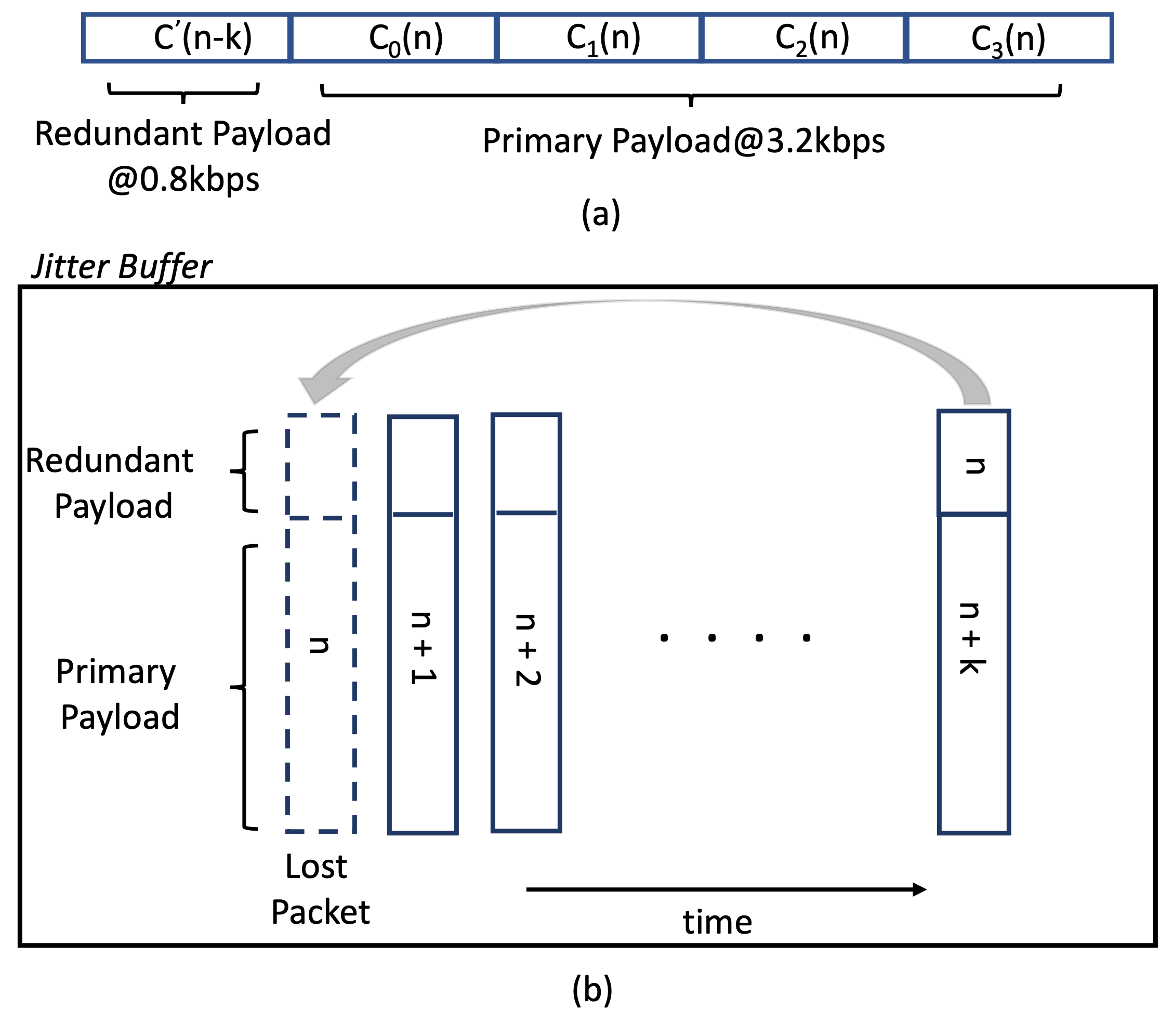}
	\caption{(a) The composition of each payload with codebooks $C_s$ of primary frame, where $s$ denotes residual stage, and codebook $C'$ of redundant frame. (b) An illustration of the FEC at jitter buffer where lost payload is concealed with redundant information, k is the frame index offset between primary and the redundant information. }
	\label{fig:fec}
	\vspace*{-10pt}
\end{figure}

Our FEC solution consists of sending the low bitrate distilled codebook index as a redundant data. The FEC method works in conjunction with JBM~\cite{EVS_JBM} and is made possible because of availability of future frames in the jitter buffer. As shown in Fig.~\ref{fig:fec}, each $(n+k)^\text{th}$ packet contains a primary data along with redundant information of the past $n^\text{th}$ frame. When $n^\text{th}$ packet is marked as lost, the corresponding redundant information at $(n+k)^\text{th}$ frame can be employed for correction. The parameter $k$ denotes the separation in terms of number of frame between the primary and the redundant payload and is called "FEC offset". The optimal value of $k$ is dictated by the length of the jitter buffer and the network conditions. For optimal transmission, the offset can be made adaptive and optimized depending on the network conditions and is usually sent along with the packet. FEC is particularly advantageous as it can provide an optimal guided correction as well as assist the PLC network to reduce its prediction error in case of burst losses. Depending on the quality or robustness requirements, the method is extremely flexible and can be used in multiple ways: a distilled codebook of larger bitrate can provide better quality of concealment whereas multiple redundant information from different offsets can be appended with primary information to provide better robustness against delayed and lost packets. Both design choices come at the cost of additional bitrate. For sake of simplicity, in this paper, we only explore the low-bitrate version of FEC at 0.8 kbps with a fixed offset and single redundant frame transmitted along with the primary frame. 

\subsection{Heuristics}
In case of long burst losses, performing predictive generation for the concealment may lead to inaccurate or falsified speech content. As countermeasure, we adjust the concealment based on the type of last received frame. If the last available frame is voiced we conceal the frames for 100 ms of burst and in case of an unvoiced frame we stop the concealment after 60 ms. There is no direct classification and segmentation of speech at the decoder, but rather a mapping between distilled codebook indices and the voiced and unvoiced classes. The mapping is performed off-line by simply observing the statistics of code-vectors on different speech segments. It was found that the codebook indices can easily be clustered into a silence, a voiced and an unvoiced class. We maintain a list of indices for different classes and use it to classify the frames during inference. 


\section{Experimental Setup}

\subsection{Training \& Inference}
The training of the PLC model requires a pre-trained NESC model. The sum of first two code-vectors of trained codebooks of NESC is used as input for distillation. The new code-vectors are updated with exponential moving average of the input with decay of 0.99 and MSE loss between input and output code-vectors is used for training. The distillation only requires few epochs for convergence after which the PLC model is trained with a teacher forcing method. A sequence of latent vectors corresponding to two seconds of audio data is presented to the PLC network that predicts the indices of subsequent code-vectors in the distilled codebook. Negative log-likelihood loss is used for training which is done for 420k iterations using ADAM optimizer at a learning rate of 0.0001 with batch size of 128.

During inference, the received past primary information are re-quantized using the distilled codebook. We maintain a history buffer containing the last seven frames of distilled code-vectors which is then used as an input to the PLC network in case of packet loss. At the softmax layer, we select the index with maximum probability for concealment. In case of burst losses, the history buffer is injected with predicted code-vectors such that the PLC model can run auto-regressively. For in-band FEC, an offset of six frames was chosen and we use the same JBM as used in EVS~\cite{EVS_JBM}. Thus, the overall bitrate of NESC with FEC is 4 kbps. For both cases of PLC and in-band FEC, the predicted or the redundant distilled code-vectors is used as inputs to NESC decoder which then produces corresponding speech signals. 

\subsection{Datasets \& Loss Traces}
The dataset used for training the distilled codebook and the PLC model was total of 280 hours of speech from LibriTTS  dataset~\cite{zen2019libritts} and VCTK dataset~\cite{VCTK} at 16 kHz. The speech signal was also augmented with background noise from the DNS Challenge dataset~\cite{dns_dataset} and reverberation from the SLR28 dataset~\cite{slr_dataset}.

We used two different datasets for evaluation: The blind dataset used from the Deep-PLC challenge 2022~\cite{dplc_2022} was used for objective evaluation. It contains 966 recordings along with corresponding loss traces. The loss traces are divided into three subsets according to the corresponding burst lengths of 120 ms, 320 ms and 1000 ms.  For subjective evaluation, we select 24 items from the NTT-AT~\cite{nttdb:2012} dataset equally balanced between two female and two male speakers. We use two delay-loss profiles with the highest error rates from ~\cite{CAprofile}. It is obtained from real-world call logs of RTP packet collected in varying network conditions. Unlike the previous traces, it not only contains indication of lost packets but also marks the packet arrival time required by the JBM. In addition to lost packets, the JBM can declare a packet as lost if the arrival time of the packet exceeds the buffer capacity. All loss traces are provided for 20 ms frame size whereas NESC operates at 10 ms frame size. In order to achieve synchronized frame losses for comparison across all baseline methods, we pack two frames of NESC in a single packet and simulate packet loss with given traces.

\subsection{Evaluation}
For evaluation, we carry out both objective and subjective assessment. For objective assessment, we use POLQA v3~\cite{POLQA}, PLCMOS~\cite{plcmos} and VISQOL v3~\cite{visqol}. PLCMOS is exclusively designed to estimate the Mean Opinion Score (MOS) when some parts of speech signals are concealed for missing packets whereas VISQOL is designed to evaluate the overall quality of speech signals. Both the methods try to predict the MOS of subjective evaluations. They are probably better suited for our evaluation because the signals generated by neural codecs or other generative networks do not necessarily preserve the waveform and hence are penalized on other audio-feature based objective evaluation like POLQA. For subjective assessment, we conduct a P.808 ACR listening test~\cite{p808} using the Amazon Mechanical Turk service involving 24 participants and accumulating 96 opinion scores per condition. 

\subsection{Baseline Methods}
For evaluation, the proposed model is compared to the following baseline models:
\begin{itemize}
	
	\item {} For comparison with conventional methods, we use EVS codec at 5.9 kbps, 8 kbps and 13.2 kbps. It performs PLC at all bitrates, but only at 13.2 kbps, it supports FEC in the Channel-Aware (CA) mode and is used to compare our FEC solution. The 5.9 kbps codec is used for objective evaluation and the other bitrates are used for P.808 because of their comparable quality with NESC.
	\item {} For PLC with a DNN-based solution, we select the LPCNet-based PLC method. The model is open source and we utilize the available pre-trained models in causal mode. Since it works as a post-processor, for comparison and in order to evaluate only the performance over the lost frames, the original signal is decoded by NESC and the concealment is performed over it. To keep the implementation simple, we do not use JBM with this method but use the traces obtained from the JBM with NESC to create loss traces.
	\item {} The naive baseline is the zero-filled NESC output where we simply select the codebook index corresponding to a silent frame for lost packets and decode it.
	
\end{itemize}

\begin{table}[t]
	\begin{center}
		\begin{adjustbox}{width=1\linewidth}
			\begin{tabular}{c  c  c  c  c}
				\toprule
				\textbf{Channels} & \textbf{Codecs} & \textbf{VISQOL} & \textbf{POLQA}  & \textbf{PLCMOS} \\ 
				\midrule
				\multirow{ 2}{*}{Clean} & EVS 5.9 kbps & 3.0699  & 3.4339 & 4.2902\\
				& NESC 3.2 kbps & 2.8849 &2.7921  & 4.3934  \\
				
				\midrule
				\multirow{ 6}{*}{Error-Prone} & Zero-filled NESC & 2.4339 & 1.8103  & 3.1121 \\
				
				&EVS 5.9 kbps & 2.5658 &  \textbf{2.1468}  & 3.3754 \\
				&NESC 3.2 kbps & \multirow{ 2}{*}{2.5208} & \multirow{ 2}{*}{1.9833}  & \multirow{ 2}{*}{\textbf{3.9139}} \\
				& \textit{ \footnotesize + proposed PLC} &&&\\
				
				&NESC 3.2 kbps & \multirow{ 2}{*}{\textbf{2.6122}} & \multirow{ 2}{*}{2.1389}  & \multirow{ 2}{*}{3.5953} \\
				&  \textit{ \footnotesize + LPCNet-PLC} &&&\\
				\toprule
			\end{tabular}
		\end{adjustbox}
		\caption{Average VISQOL, POLQA and PLCMOS scores for comparison of proposed solution with the discussed baselines. Higher scores correspond to better quality.}
		\label{tab:obj_scores} 
	\end{center}
\end{table}
\vspace*{-10pt}
\begin{figure}[t]
	\vspace*{-10pt}
	\centering
	\includegraphics[width=\linewidth]{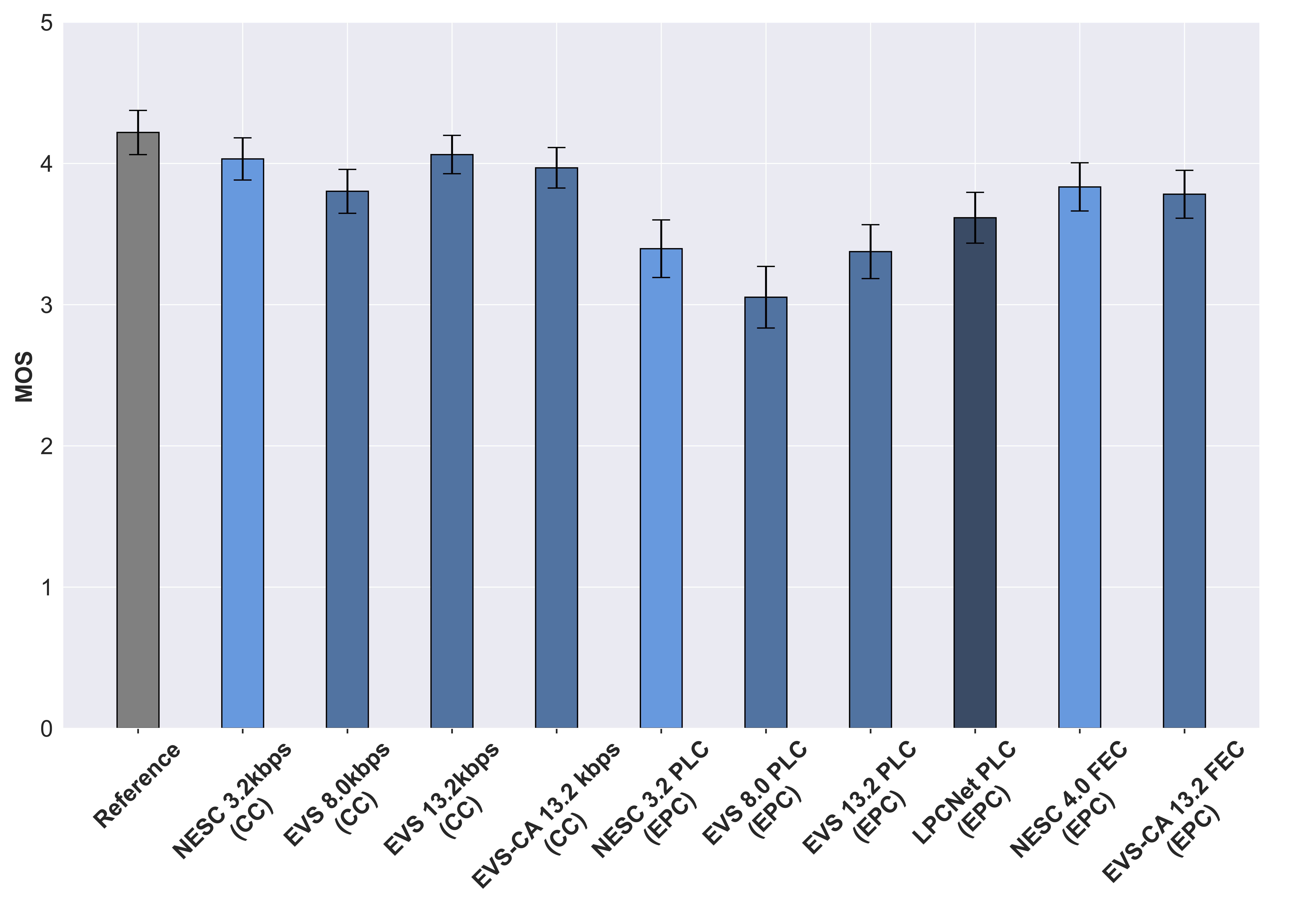}
	\caption{Listening test results for loss traces in Profile-1, with 95\% confidence intervals. CC and EPC stands for Clean Channel and Error-prone channel respectively. }
	\label{fig:p808_p1}
	\vspace*{-10pt}
\end{figure}

\begin{figure}[t]
	\centering
	\includegraphics[width=\linewidth]{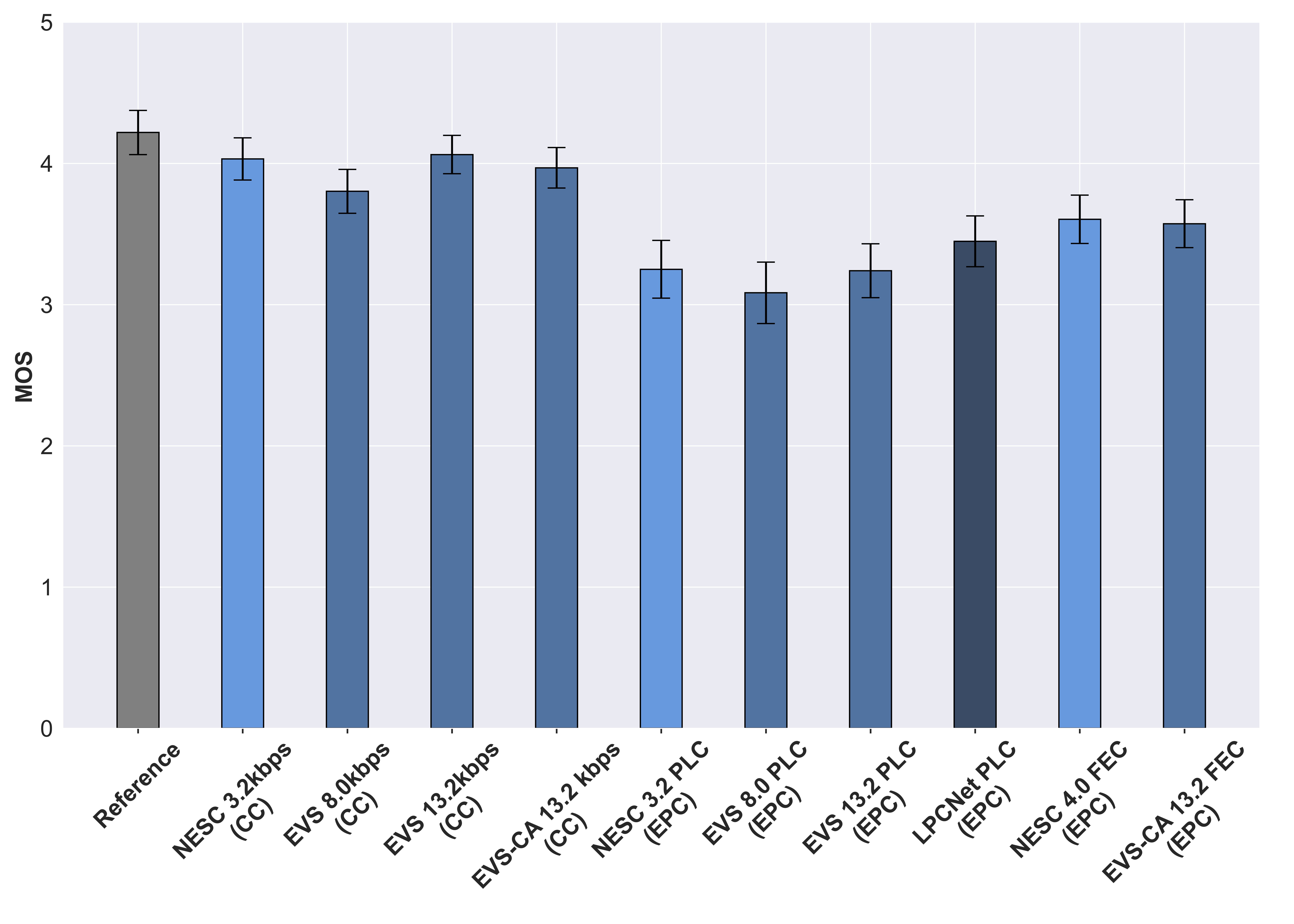}
	\caption{Listening test results for loss traces in Profile-2, with 95\% confidence intervals. CC and EPC stands for Clean Channel and Error-prone channel respectively.}
	\label{fig:p808_p2}
	\vspace*{-10pt}
\end{figure}

\section{ Results and Discussion}
In Table~\ref{tab:obj_scores}, we present the average objective scores obtained for various methods. In all the measures, the zero-filled NESC shows the lowest scores which illustrates the distortion caused by packet losses without concealment. The quality measures POLQA and VISQOL rates the concealment provided by EVS-PLC and LPCNet-PLC as the highest whereas PLCMOS rates our proposed solution as the best. This difference in quality measures can be attributed to the fact that our proposed PLC solution operates at the lowest bitrate with a very coarsely quantized level of the latent. The quality of the speech signal generated for concealment in our proposed solution is somehow equivalent to operating NESC at 0.8 kbps. On the other hand, the LPCNet-PLC operates at the output signal and uses a dedicated additional neural vocoder with calculated un-quantized feature for generation. However, given the low computational overhead that the model entails, NESC PLC shows an interesting trade-off between performance and complexity, and provides substantial benefit over the zero-filled baseline. We do not present the objective assessment of the FEC solution because of the paucity of loss traces with packet arrival time.

The subjective scores are split into two parts based on the loss profile used and are shown in Figure~\ref{fig:p808_p1} and Figure~\ref{fig:p808_p2}. Profile 1 and 2 simulate packet loss rates of about 8\% and 10\%, respectively. In comparison to Profile-1, Profile-2 contains burst losses of higher lengths and simulates higher error rates due to delays in packet arrival time. We include EVS at multiple bitrates to understand the granularity of quality. In clean channel, without packet losses, the listeners reported NESC at 3.2 kbps to have similar quality as EVS 13.2 kbps. EVS in CA mode shows a slight drop in quality because the CA mode reserves 0 to 3.6 kbps of bitrate per frame for FEC. In error-prone channel, the NESC PLC performs at the same level as EVS 13.2 kbps and is slightly below the LPCNet-PLC method.

On the other hand, the listening test results show the effectiveness of our proposed low-bitrate FEC solution. It is at least on par with the CA mode of EVS and is better than compared stand-alone PLC solutions. The results show that, for neural codecs, in-band FEC in conjunction with PLC is capable of providing very high-quality error resilience\footnote{Check our demo samples at: \href{https://fhgspco.github.io/nesc_plc_fec/}{\url{https://fhgspco.github.io/nesc_plc_fec/}}}. 

In terms of complexity, our proposed PLC model contains 0.6 million parameters and has a complexity of 65.5 MFLOPS. At 10\% loss of packets, the integrated solution of NESC with PLC shows 3\% decrease in real-time factor compared to stand-alone NESC. The measurement was done on a single thread of an Intel(R) Core(TM) i7-6700 CPU at 3.40GHz.

\section{Conclusion}
In this paper, we provide error resilient tools for end-to-end neural coders, taking into account real constraints on both the added complexity and the network characteristics. A low-complexity PLC model is proposed, which operates directly in the latent domain and exploits vector quantization, taking advantage of the generative capability of the decoder. In addition, to limit the need for concealment, we proposed the use of in-band FEC, to correct and decode lost packets using the redundant information transmitted at an additional bitrate of 0.8 kbps. In future work, we intend to extend this method to other end-to-end neural codecs and evaluate its effectiveness. We also plan to explore the use of neural networks for in-band FEC.

\bibliographystyle{IEEEtran}
\bibliography{mybib}

\end{document}